\documentclass[twocolumn,aps,prl,groupedaddress,superscriptaddress,amsmath,amssymb,showpacs,noeprint,english]{revtex4-2}

\usepackage{graphicx}
\usepackage{dcolumn}
\usepackage{bm}
\usepackage[T1]{fontenc}
\usepackage{mathrsfs}
\usepackage{amsbsy}
\usepackage{amstext}
\usepackage{amssymb}
\usepackage{setspace}
\usepackage{esint}
\usepackage{xcolor,colortbl}
\usepackage[colorlinks,citecolor=blue,urlcolor=blue,linkcolor=blue,hypertexnames=true]{hyperref} 
\usepackage{babel}

\usepackage{tikz}

\newcommand*\fullcirc{\tikz\fill (0,0) circle (0.6ex);}

\begin{document}

\title{Model-Independent Quantum Phases Classifier}

\author{F. Mahlow}
\email{f.mahlow@unesp.br}
\affiliation{Faculty of Sciences, UNESP - S{\~a}o Paulo State University, 17033-360 Bauru, S{\~a}o Paulo, Brazil}
	
\author{F. S. Luiz}
\affiliation{Faculty of Sciences, UNESP - S{\~a}o Paulo State University, 17033-360 Bauru, S{\~a}o Paulo, Brazil}

\author{A. L. Malvezzi}
\affiliation{Faculty of Sciences, UNESP - S{\~a}o Paulo State University, 17033-360 Bauru, S{\~a}o Paulo, Brazil}

\author{F. F. Fanchini}
\affiliation{Faculty of Sciences, UNESP - S{\~a}o Paulo State University, 17033-360 Bauru, S{\~a}o Paulo, Brazil}

\date{\today}
\begin{abstract}
Machine learning has revolutionized many fields of science and technology. Through the $k$-Nearest Neighbors algorithm, we develop a model-independent classifier, where the algorithm can classify phases of a model to which it has never had access. For this, we study three distinct spin-$1$ models with some common phases: the XXZ chains with uniaxial single-ion-type anisotropy, the bound alternating XXZ chains, and the bilinear biquadratic chain. We show that, with high probability, algorithms trained with two of these models can determine common phases with the third. It is the first step toward a universal classifier, where an algorithm is able to detect any phase with no knowledge about the Hamiltonian, only knowing partial information about the quantum state.
\end{abstract}
\maketitle


\textit {Introduction ---} The interest on quantum phases and phase transitions has been recently renewed due to new physics uncovered by experiments on cuprate superconductors, heavy fermion materials, organic conductors, and other strongly correlated materials \cite{Paschen2021}. From a theoretical perspective, low-dimensional quantum lattice models can capture many aspects of these new phenomena \cite{Dagotto1994,Dzero2016}. Therefore, determining the quantum (i.e., groundstate) phase diagrams of these models is an important challenge in condensed matter and statistical physics \cite{Sachdev2011}.

Concerning quantum phases classification, many alternatives have been highlighted as promising \cite{Hastings2013,Schuch2011,Chen2011,Carleo2017,Huang2021} with special emphasis on machine learning (ML). The main difference between ML and other statistical models is the fact that 
an algorithm can improve its performance, that is, learn, without the need for such explicit programming \cite{Schuld2018, Mehta2019}. ML is a form of applied statistics, where computers use data (usually in large quantities) to estimate functions with a high degree of complexity, which can be used to make predictions and observe patterns in these data sets \cite{Mitchell1997,Goodfellow2016}. ML has been widely used in the physical sciences, including cosmology \cite{Kamdar2015, Kamdar2016, Lochner2016}, quantum information \cite{Torlai2018, Canabarro2019a, Iten2020}, many-body physics \cite{Carleo2017}, and also to classify quantum phases and detect their transitions \cite{Carrasquilla2017, Dong2019, Shiina2020, Rem2019, Broecker2017, Canabarro2019}.

In this letter, we analyze the correlation between spins in a closed chain for three distinct spin-1 models. We show that these correlations hold information about the phases of these distinct models, and there is considerable overlap between phase information of different models. This explain how ML algorithms, training a known model, are capable to detect some phases of another unknown model. To illustrate we use a machine learning classifier, fed with correlation and phases labels of two known models, to detect the phases of a third unknown model. We show that the prediction is succeed when the overlap of the information about the different phases is minimal. Also, we show it is possible to apply a transformation in the dataset of the correlations, which allows minimizing the overlap of the information about the different phases, making the ML predictor more accurate.

\textit{The Models ---} We use three well-known distinct spin-1 models, where the phase diagram of these models and their central features are well established in the literature \cite{Kitazawa1996,Chen2003,Luchli2006}. The first mode, we present is the \textit{(i) XXZ chains with uniaxial single-ion-type anisotropy}, whose Hamiltonian is given by:
\begin{equation}
{\cal H}_{1} = \sum_{l=1}^{N} [J({S_l^x}{S_{l+1}^x}+{S_l^y}{S_{l+1}^y})+J_z{S_l^z}{S_{l+1}^z}]+ D\sum_{l=1}^{N} {S_l^z}^2,
\label{h1}
\end{equation}
where $S_l$ is a spin-$1$ operator acting on site $l$ of a one-dimensional lattice (chain) with $N$ sites, $D$ represents uniaxial single ion anisotropy, and $J(=1), J_z$ are spin couplings. For Hamiltonian Eq. \ref{h1}, the phase diagram consists of six distinct phases, namely, Haldane, Large D, XY1, XY2, Ferromagnetic, and a N{\'e}el, and several different transitions can occur \cite{Chen2003}. The next model is the \textit{(ii) bond-alternating XXZ chain}, whose Hamiltonian is given by:
\begin{equation}
{\cal H}_2 = \sum_{l=1}^{N} [1-\delta(-1)^l][{S_l^x}{S_{l+1}^x} + {S_l^y}{S_{l+1}^y} + \Delta {S_l^z}{S_{l+1}^z}],
\label{h2}
\end{equation}
where $\Delta$ is the strength of the Ising-type anisotropy that originates from the spin-orbit interaction in magnetic materials and $\delta$ is the alternation of the bond that describes dimerization. The phase diagram of the model Eq. \ref{h2} shows the Ferromagnetic phase, XY1, N{\'e}el, Haldane, and Dimerized \cite{Kitazawa1996}. Finally, the last model analyzed was the \textit{(iii) bilinear biquadratic chain}, whose Hamiltonian is given by:
\begin{equation}
{\cal H}_3 = \sum_{l=1}^{N} [\cos\theta(S_l.S_{l+1})+ \sin\theta (S_l.S_{l+1})^2],
\label{h3}
\end{equation}
where $\theta \in [0, 2 \pi)$ quantifies the amount of coupling between the nearest neighboring spins. The model Eq. \ref{h3} presents four phases, namely, Haldane, Trimerized, Ferromagnetic, and Dimerized  \cite{Luchli2006}. 

To illustrate and summarize the phases and the common phases contained among these models, we prepare Table {\ref{tab1}}. As we can be noted, all phases of ${\cal H}_2$ are contained in the combined phases of ${\cal H}_1$ and ${\cal H}_3$, being the only phase diagram in which this occurs. The union of ${\cal H}_1$ and ${\cal H}_2$ contains three of four phases of ${\cal H}_3$, and the union of ${\cal H}_2$ and ${\cal H}_3$ contains four of six phases of ${\cal H}_1$. In total, we have five phases that are shared by at least two models (Haldane, Néel, Ferromagnetic, XY1, and Dimer) and three phases unique to a single model (Large-D, XY$2$, and Trimer). 



We assume that even when different models appear on the phase diagram, a given phase has a trademark that is model independent. We propose here that the spin correlations can capture such a signature. To test this hypothesis, we will analyze several correlations between the spins in the chain.

\newcolumntype{P}[1]{>{\centering\arraybackslash}p{#1}}
{\centering
 \begin{table}[h]
    \begin{tabular}{c|P{1.5cm}P{1.5cm}P{1.5cm}}
    \hline
      \textbf{Quantum Phase} & {${\cal H}_1$} & {${\cal H}_2$} & {${\cal H}_3$} \\ 
      \hline
      Haldane & $\color{red}{\fullcirc}_{(288)}$ & $\color{red}{\fullcirc}_{(473)}$ & $\color{red}{\fullcirc}_{(1150)}$ \\
      N{\'e}el & $\color{olive}{\fullcirc}_{(2081)}$ & $\color{olive}{\fullcirc}_{(1284)}$ &  \\
      Ferromagnetic & $\color{blue}{\fullcirc}_{(2168)}$ & $\color{blue}{\fullcirc}_{(800)}$ & $\color{blue}{\fullcirc}_{(1725)}$ \\
      Large-D & $\color{yellow}{\fullcirc}_{(1623)}$ &  &  \\
      XY1 & $\color{cyan}{\fullcirc}_{(207)}$ & $\color{cyan}{\fullcirc}_{(1028)}$ &  \\
      XY2 & $\fullcirc_{(33)}$ &  &  \\
      Dimer &  & $\color{green}{\fullcirc}_{(2815)}$ & $\color{green}{\fullcirc}_{(1150)}$ \\
      Trimer &  &  & $\color{purple}{\fullcirc}_{(575)}$ \\\hline
    \end{tabular}\caption {Phases contained in the diagrams corresponding to the three Hamiltonians analyzed. The filled circles mean that the quantum phase is present in the model. The numbers subscribed in the circles represent the amount of data calculated for each phase of each model.}\label{tab1}\end{table}\par}


\textit{Data structure ---} The correlations between neighbors in the closed spin chain are given by the expected values of the following observables: $\langle S_{1}^{k}, S_{i}^{k}\rangle$ and $\langle\prod_{j} S_{j}^{k}\rangle $,  with $k=\{ x,y,z\}$, $ i = [1, N/2+1]$, and $j=[1,N]$. Furthermore, $\langle S_{1}^{k}  S_{i}^{k}\rangle = \langle\lambda_{0}| S_{1}^{k},  S_{i}^{k}|\lambda_{0}\rangle$, are the expected values of the correlation for the lowest energy state of the Hamiltonian and we take the number of spins $N=12$ \cite{Canabarro2019}. Notice that since the chain is closed and, consequently, the chain properties are cyclic, any non-redundant correlation between two spins is obtained for $\langle S_{1}^{k}, S_{i}^{k}\rangle$ with $ j = [1, N/2+1]$.

To generate the correlation dataset, we considered thousands of different values for the parameters of the Hamiltonians ${\cal H}_1$, ${\cal H}_2$, and ${\cal H}_3$. To ${\cal H}_1$, Eq. \ref{h1}, we range the parameters $J_{z}$ and $D$ in the interval $[-4,4]$ with step size of $0.1$, this generates a dataset with $6400$ data points. For the Hamiltonian ${\cal H}_2$, Eq. \ref{h2}, we range the parameters, $\Delta$ in an interval $[0,1]$,  and the parameters $\delta$ in an interval $[-1.5,2.5]$, with step sizes of 0.005 and 0.0125, respectively, again this generate a dataset with $6400$ data points. Finally, for the Hamiltonian ${\cal H}_3$, Eq. \ref{h3},  we set the range of parameter $\theta$ in the interval $[0,2\pi]$, with the step size of $4.35\times 10^{-4}$, which results in a dataset with $4600$ data points. The labels of the phases for the Hamiltonians ${\cal H}_1$, ${\cal H}_2$, and ${\cal H}_3$, are obtained from the literature \cite{Chen2003}, \cite{Kitazawa1996}, and \cite{Luchli2006}, respectively. 

\begin{figure*}
\centering
\includegraphics[width=\textwidth]{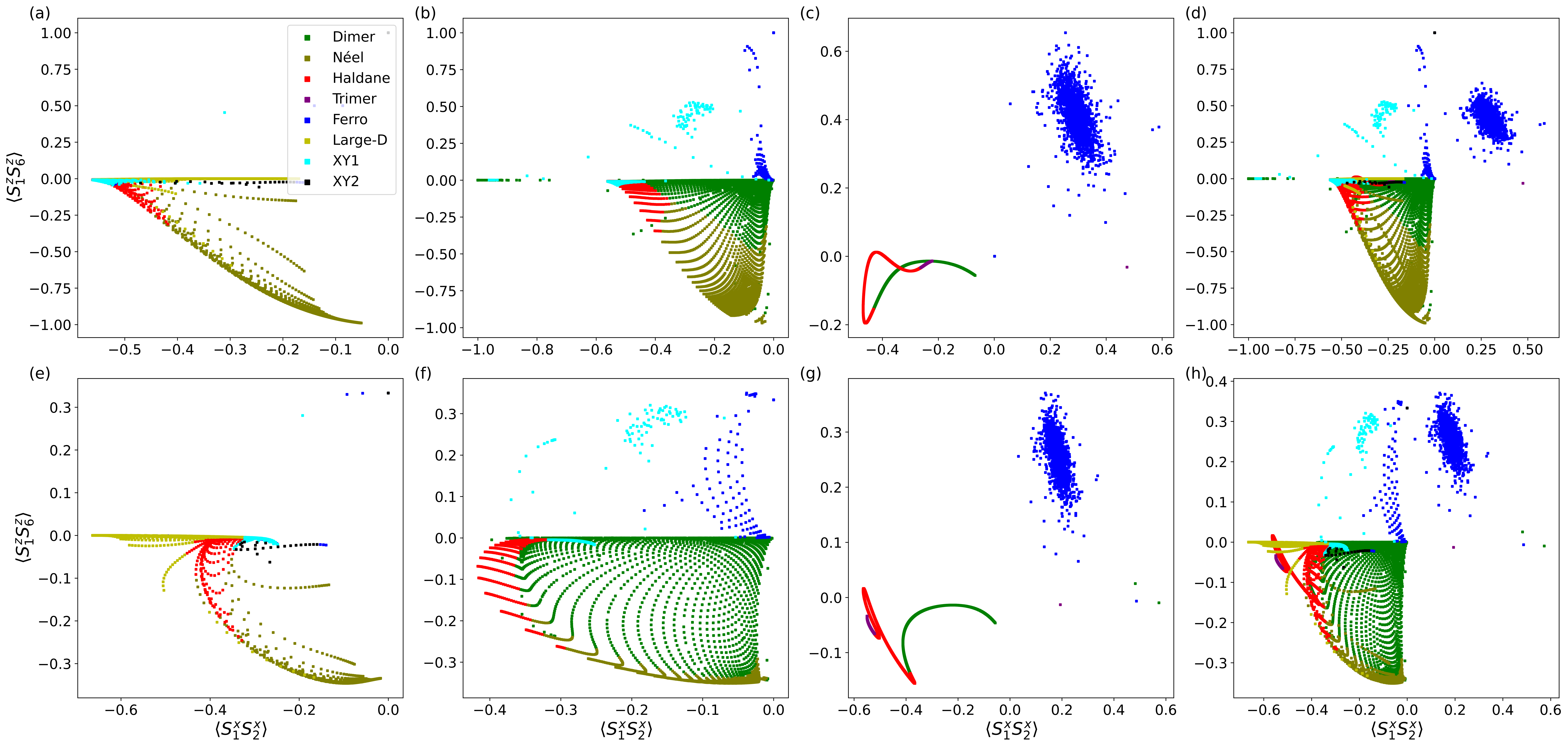}
\caption{Correlation $\langle S^{x}_{1}S^{x}_{2}\rangle$ vs correlation $\langle S^{z}_{1}S^{z}_{6}\rangle$, the color represents the phase of the corresponding data point. And the figures (a) and (e) refer to the system described by ${\cal H}_{1}$, (b) and (f) refers to ${\cal H}_{2}$, (c) and (g) refers to ${\cal H}_{3}$, (d) and (h) are the data from the three systems. The figures (a), (b), (c) and (d) are the raw data, and (e), (f), (g), (h) are the data after performing the transformation.} 
\label{fig_data}
\end{figure*}

With the dataset of the three models, we could visualize the relation of the correlations with the quantum phases. Since we intend to use a classifier algorithm, the idea is to separate, in the multidimensional space, distinct phases in distinct positions. For $12$ sites, for example, using the set of correlations described above, we have $24$ distinct correlations (a space with $24$ dimensions) and to visualize this amount of information in a 3D space is complicated. Nevertheless, a clue as to what happens in the correlation space can be obtained by considering only a pair of correlations. To illustrate this, in  Fig. (\ref{fig_data}) we choose a specific pair of correlations to plot, namely, $\langle S^{x}_{1}S^{x}_{2}\rangle$ and $\langle S^{z}_{1}S^{z}_{6}\rangle$, looking for the best graphical representation of what happens in the correlation space. Note that {$\langle S^{x}_{1}S^{x}_{2}\rangle$ and $\langle S^{z}_{1}S^{z}_{6}\rangle$} are calculated for all dataset, where we consider different values for the parameters of the Hamiltonians ${\cal H}_1$, ${\cal H}_2$, and ${\cal H}_3$.


One important aspect to consider when working with machine learning is a dataset transformation. In many cases, appropriate transformations can separate the classes in the feature space, which facilitates the classification process. In Fig. (\ref{fig_data}-a) to Fig. (\ref{fig_data}-d) we plot the raw data, and in Fig. (\ref{fig_data}-e) to Fig. (\ref{fig_data}-h) we plot the raw data after scaling the dataset, for each model, to have a unit norm. This is a well-known renormalization procedure called spatial sign preprocessing \cite{Serneels}.  Analyzing Fig. (\ref{fig_data}), we see that even with the information contained in a single pair of correlations, we are able to separate, with high distinctness, states with different phases. Indeed, there is some overlap of this information for different phases, especially when considering several different models, Fig. (\ref{fig_data}-d) and Fig. (\ref{fig_data}-h). Moreover, for the normalized data, Fig. (\ref{fig_data}-e) to Fig. (\ref{fig_data}-h), the overlap of information about the phases decreases for all models, almost disappearing for ${\cal H}_3$, Fig. (\ref{fig_data}-g). 

Despite the improvement in reducing the overlap, Fig. (\ref{fig_data}-h) shows that there are still overlaps of information from the Haldane phase with Large-D, from the XY1, XY2, Néel, and Haldane phases with the Dimmer phase. 
However, it is important to note that we are illustrating the procedure considering only two features, which is insufficient to separate all phases given the complexity of the problem. In fact, the use of the 24 features, as well as studying and applying other transformations to separate the phases in the correlation space, becomes crucial and is the focus of this work. Our main goal is to classify the phases of an unknown model, given the distribution of these phases in the correlation space of other known models. For this purpose, we need an algorithm that, given a new quantum state and, consequently, a position in the correlation space, labels it with the corresponding phase based on the information from other known states, located in the vicinity of the new one in the correlation space. As we  show below, even a simple algorithm is capable of performing this task with high accuracy.


\textit{Machine Learning ---} In this manuscript, we use an supervised algorithm, i.e. where the learning process occurs through labeled data. To this task, one of the the simplest machine learning algorithms is the $k$-Nearest Neighbors classifier ($k$-NN) \cite {Andreas2016}.
Despite its simplicity, it presented a good result in our classification problem, which can be explained by the way the $k$-NN works. When a data point of the unknown model is inserted,  the algorithm calculated the Euclidean distance (in this work, but the metric can be changed) of this unknown model data point to the $k$-nearest neighbors. The unknown model data point is classified by a plurality vote of its $k$ neighbors, with the unknown model data point being assigned to the class most common among them. For example, if $k = 1$, then the object is simply assigned to the class of that single nearest neighbor \cite{Andreas2016,scikit-learn,Evelyn1951}. 

In this work, we use $k=30$ and assume that all neighbors have the same voting weight (it could be assumed that the closer, the greater the voting weight). The values were obtained by performing the hypermetrization of the parameters \cite{scikit-learn}. Analyzing Fig. (\ref{fig_data}) and understanding how $k$-NN works, it quite plausible that a high accuracy universal classifier can be obtained. Naturally, it is necessary to use an appropriate transformation that reduces the overlap between the different phases in the correlation space. Once the transformation decreases the overlap between different phases (such as that made in Fig. (\ref{fig_data}-e) to Fig. (\ref{fig_data}-h)), it increases the accuracy of the $k$-NN algorithm.


\textit{Results ---} To test the accuracy of the $k$-NN in transferring learning from the known models to the unknown model, we begin by training the algorithm to classify the data points of ${\cal H}_1$. For this case, we train the $k$-NN with data from models ${\cal H}_2$ and ${\cal H}_3$ (it means that the $k$-NN algorithm knows the dataset of ${\cal H}_2$ and ${\cal H}_3$ and their respective phase labels), and predict the phases of ${\cal H}_1$. In sequence, we do the same for ${\cal H}_2$ and ${\cal H}_3$ using the data from models ${\cal H}_1$ and ${\cal H}_3$, and ${\cal H}_1$ and ${\cal H}_2$, respectively.
It is worth noting, however, that there are phases in ${\cal H}_1$ (Large-D and XY2) that do not exist in ${\cal H}_2$ and ${\cal H}_3$, and a phase in ${\cal H}_2$ (Trimer) that does not exist in ${\cal H}_1$ and ${\cal H}_3$. Clearly, there is no way to learn from the data that were not provided \cite{Huang2021,Huang2021b}, for that reason, when we predict the phase of ${\cal H}_1$, we remove the data for the Large-D and XY2 phases. Furthermore, when we predict the phases of ${\cal H}_2$ we remove the Trimer phase from the predicted dataset, which ensures that unknown phases do not interfere with the accuracy of the $k$-NN. 


\begin{figure*}
\centering
\includegraphics[width=\textwidth]{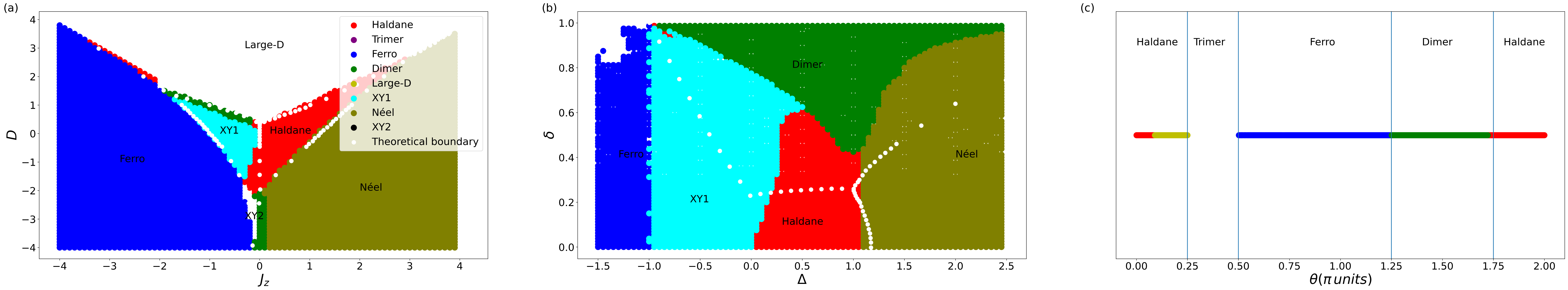}
\caption{Figure (a) is the phase prediction of model ${\cal H}_1$ using the learning of models ${\cal H}_2$ and ${\cal H}_3$. Figure (b) is the phase prediction of model ${\cal H}_2$ using the learning of models ${\cal H}_1$ and ${\cal H}_3$, and figure (c) is the phase prediction of model ${\cal H}_3$ using the learning of models ${\cal H}_1$ and ${\cal H}_2$. The white dots in panels (a) and (b), and light blue line in panel (c) are the theoretical boundaries of phases given in the literature, \cite{Chen2003, Kitazawa1996, Luchli2006}, for models ${\cal H}_1$, ${\cal H}_2$, and ${\cal H}_3$, respectively. The name in black represents the real phase in that place of the model, and the color represents the predicted phase.} 
\label{fig_prev}
\end{figure*}

In our analysis, we transform our dataset following a renormalization procedure called spatial sign preprocessing, where the dataset is scaled to have a unit norm \cite{Serneels}. As stated before, each ground state and its respective labeled phase, are represented by the correlations between a pair of spins, $\langle S_{1}^{k},S_{i}^{k}\rangle$, which provide 7 features for each variable $k=\{x,y,z\}$, and global correlations $\langle\prod_{j} S_{j}^{k}\rangle$, which provide 3 more features. As we note in Fig. (\ref{fig_data}), where we use only 2 correlations to illustrate our strategy, even after the dataset transformation, some overlap between phases is still present. For example, in Fig. (\ref{fig_data}-f), it is easy to notice the overlap of the XY1 phase with Haldane and Dimer phase, and the overlap between Néel and Dimmer phases. So, the question emerges: is the inclusion of all features able to separate the phases in the correlation space? To answer this question, Fig. (\ref{fig_prev}) shows the phase prediction of the $k$-NN algorithm, where in Fig. (\ref{fig_prev}-a) we estimate the phase diagram of ${\cal H}_1$ given the data from ${\cal H}_2$ and ${\cal H}_3$, in Fig. (\ref{fig_prev}-b) we estimate the phase diagram of model ${\cal H}_2$ given the data from ${\cal H}_1$ and ${\cal H}_3$, and analogous in Fig. (\ref{fig_prev}-c) where we estimate the phases of the ${\cal H}_3$ given the data from ${\cal H}_1$ and ${\cal H}_2$.

The prediction of the phase diagram of ${\cal H}_1$, presented in Fig. (\ref{fig_prev}-a), was incredibly successful with an accuracy of $96.31\%$. All phases are in the correct locations with few mistakes in the phase transitions. As in our two-dimensional illustrative example, the Haldane phase invades the XY1 phase space, and the confusion between the Néel and Dimmer phase persists. Nevertheless, analyzing Fig. (\ref{fig_data}-f), it is important to note that the $k$-NN correctly predicts the XY1 phase, even thought this phase is completely overlapped with Dimer and Haldane phase in the two-dimensional case.
When considering ${\cal H}_2$, we observe that the algorithm commits mistakes in the separations of the phases, which made its accuracy the lowest $72.72\%$. Despite this, one aspect needs to be emphasized. The training dataset contains information for ${\cal H}_1$ and ${\cal H}_3$, which includes all 8 distinct phases.  Nevertheless, the Large-D, XY2, and Trimer phases were not indicated by the $k$-NN algorithm for all data of ${\cal H}_2$, since the algorithm detects all phases correctly, only making mistakes around the boundaries.


Finally, the prediction for model ${\cal H}_3$ is presented in Fig. (\ref{fig_data}-c), where we use the ${\cal H}_1$ and ${\cal H}_2$ models for learning. In one hand, the confusion between the Haldane phase and the Dimer phase presented in Fig. (\ref{fig_data}-f) was removed when all 24 features were included. However, it incorrectly classifies the Haldane phase, mixing it with the Large-D, a phase that is not even present in the model. Nevertheless, even making this mistake, the algorithm achieve an accuracy of $89.17\%$.
Although the good results, different strategies can be used to increase the prediction accuracy. One is to add new models to the predictor dataset, as adding new information to the $k$-NN would help avoid incorrect phase prediction. The second is to find a transformation that can separate the phase information in the correlation space. Finally, different machine learning algorithms can certainly be implemented to increase the accuracy. Nevertheless, these results demonstrate that the correlations between the spins provide essential information about the phases of the models. This shows that correlations obtained from known models can be used to predict phases in other different models with high accuracy.




\textit{Conclusion ---} 

We have developed a method to study the phases of unknown magnetic systems through spin correlations. We show that raw correlation data carry information about the phases, which is independent of the model. With the spin correlation information, we use a $k$-NN algorithm to predict the phases of an unknown model with high accuracy. We present a proof of concept, showing that a ML algorithm can classify unknown phases of a Hamiltonian through known phases of another Hamiltonian, creating a model-independent quantum phase classifier. We emphasize that no explicit use of the phases order parameters is made, so 
this model-independent classifier opens up the possibility of universal classifier creation, as more and more model-independent information is added to the classifier database.

\begin{acknowledgments}
F. M. and F.S.L. acknowledge support from Coordena\c{c}{\~a}o de Aperfei\c{c}oamento de Pessoal de N{\'i}vel Superior (CAPES), project number 88887.607339/2021-00 and 151435/2020-0, respectively; F.M. and F.F.F acknowledge support from Funda\c{c}{\~a}o de Amparo {\`a} Pesquisa do Estado de S{\~a}o Paulo (FAPESP), project number 2019/00700-9 and 2021/04655-8, respectively.
\end{acknowledgments}

\bibliography{bibl}

\begin{thebibliography}{33}%
\makeatletter
\providecommand \@ifxundefined [1]{%
 \@ifx{#1\undefined}
}%
\providecommand \@ifnum [1]{%
 \ifnum #1\expandafter \@firstoftwo
 \else \expandafter \@secondoftwo
 \fi
}%
\providecommand \@ifx [1]{%
 \ifx #1\expandafter \@firstoftwo
 \else \expandafter \@secondoftwo
 \fi
}%
\providecommand \natexlab [1]{#1}%
\providecommand \enquote  [1]{``#1''}%
\providecommand \bibnamefont  [1]{#1}%
\providecommand \bibfnamefont [1]{#1}%
\providecommand \citenamefont [1]{#1}%
\providecommand \href@noop [0]{\@secondoftwo}%
\providecommand \href [0]{\begingroup \@sanitize@url \@href}%
\providecommand \@href[1]{\@@startlink{#1}\@@href}%
\providecommand \@@href[1]{\endgroup#1\@@endlink}%
\providecommand \@sanitize@url [0]{\catcode `\\12\catcode `\$12\catcode
  `\&12\catcode `\#12\catcode `\^12\catcode `\_12\catcode `\%12\relax}%
\providecommand \@@startlink[1]{}%
\providecommand \@@endlink[0]{}%
\providecommand \url  [0]{\begingroup\@sanitize@url \@url }%
\providecommand \@url [1]{\endgroup\@href {#1}{\urlprefix }}%
\providecommand \urlprefix  [0]{URL }%
\providecommand \Eprint [0]{\href }%
\providecommand \doibase [0]{https://doi.org/}%
\providecommand \selectlanguage [0]{\@gobble}%
\providecommand \bibinfo  [0]{\@secondoftwo}%
\providecommand \bibfield  [0]{\@secondoftwo}%
\providecommand \translation [1]{[#1]}%
\providecommand \BibitemOpen [0]{}%
\providecommand \bibitemStop [0]{}%
\providecommand \bibitemNoStop [0]{.\EOS\space}%
\providecommand \EOS [0]{\spacefactor3000\relax}%
\providecommand \BibitemShut  [1]{\csname bibitem#1\endcsname}%
\let\auto@bib@innerbib\@empty
\bibitem [{\citenamefont {Paschen}\ and\ \citenamefont
  {Si}(2021)}]{Paschen2021}%
  \BibitemOpen
  \bibfield  {author} {\bibinfo {author} {\bibfnamefont {S.}~\bibnamefont
  {Paschen}}\ and\ \bibinfo {author} {\bibfnamefont {Q.}~\bibnamefont {Si}},\
  }\bibfield  {title} {\bibinfo {title} {Quantum phases driven by strong
  correlations},\ }\href {https://doi.org/10.1038/s42254-020-00262-6}
  {\bibfield  {journal} {\bibinfo  {journal} {Nature Review Physics}\ }\textbf
  {\bibinfo {volume} {3}},\ \bibinfo {pages} {9} (\bibinfo {year}
  {2021})}\BibitemShut {NoStop}%
\bibitem [{\citenamefont {Dagotto}(1994)}]{Dagotto1994}%
  \BibitemOpen
  \bibfield  {author} {\bibinfo {author} {\bibfnamefont {E.}~\bibnamefont
  {Dagotto}},\ }\bibfield  {title} {\bibinfo {title} {Correlated electrons in
  high-temperature superconductors},\ }\href
  {https://doi.org/10.1103/RevModPhys.66.763} {\bibfield  {journal} {\bibinfo
  {journal} {Rev. Mod. Phys.}\ }\textbf {\bibinfo {volume} {66}},\ \bibinfo
  {pages} {763} (\bibinfo {year} {1994})}\BibitemShut {NoStop}%
\bibitem [{\citenamefont {Dzero}\ \emph {et~al.}(2016)\citenamefont {Dzero},
  \citenamefont {Xia}, \citenamefont {Galitski},\ and\ \citenamefont
  {Coleman}}]{Dzero2016}%
  \BibitemOpen
  \bibfield  {author} {\bibinfo {author} {\bibfnamefont {M.}~\bibnamefont
  {Dzero}}, \bibinfo {author} {\bibfnamefont {J.}~\bibnamefont {Xia}}, \bibinfo
  {author} {\bibfnamefont {V.}~\bibnamefont {Galitski}},\ and\ \bibinfo
  {author} {\bibfnamefont {P.}~\bibnamefont {Coleman}},\ }\bibfield  {title}
  {\bibinfo {title} {Topological kondo insulators},\ }\href
  {https://doi.org/10.1146/annurev-conmatphys-031214-014749} {\bibfield
  {journal} {\bibinfo  {journal} {Annual Review of Condensed Matter Physics}\
  }\textbf {\bibinfo {volume} {7}},\ \bibinfo {pages} {249} (\bibinfo {year}
  {2016})}\BibitemShut {NoStop}%
\bibitem [{\citenamefont {Sachdev}(2011)}]{Sachdev2011}%
  \BibitemOpen
  \bibfield  {author} {\bibinfo {author} {\bibfnamefont {S.}~\bibnamefont
  {Sachdev}},\ }\href@noop {} {\emph {\bibinfo {title} {Quantum phase
  transitions}}}\ (\bibinfo  {publisher} {Cambridge University Press},\
  \bibinfo {address} {Cambridge, NY},\ \bibinfo {year} {2011})\BibitemShut
  {NoStop}%
\bibitem [{\citenamefont {Hastings}(2013)}]{Hastings2013}%
  \BibitemOpen
  \bibfield  {author} {\bibinfo {author} {\bibfnamefont {M.~B.}\ \bibnamefont
  {Hastings}},\ }\bibfield  {title} {\bibinfo {title} {Classifying quantum
  phases with the kirby torus trick},\ }\bibfield  {journal} {\bibinfo
  {journal} {Physical Review B}\ }\textbf {\bibinfo {volume} {88}},\ \href
  {https://doi.org/10.1103/physrevb.88.165114} {10.1103/physrevb.88.165114}
  (\bibinfo {year} {2013})\BibitemShut {NoStop}%
\bibitem [{\citenamefont {Schuch}\ \emph {et~al.}(2011)\citenamefont {Schuch},
  \citenamefont {Pérez-García},\ and\ \citenamefont {Cirac}}]{Schuch2011}%
  \BibitemOpen
  \bibfield  {author} {\bibinfo {author} {\bibfnamefont {N.}~\bibnamefont
  {Schuch}}, \bibinfo {author} {\bibfnamefont {D.}~\bibnamefont
  {Pérez-García}},\ and\ \bibinfo {author} {\bibfnamefont {I.}~\bibnamefont
  {Cirac}},\ }\bibfield  {title} {\bibinfo {title} {Classifying quantum phases
  using matrix product states and projected entangled pair states},\ }\bibfield
   {journal} {\bibinfo  {journal} {Physical Review B}\ }\textbf {\bibinfo
  {volume} {84}},\ \href {https://doi.org/10.1103/physrevb.84.165139}
  {10.1103/physrevb.84.165139} (\bibinfo {year} {2011})\BibitemShut {NoStop}%
\bibitem [{\citenamefont {Chen}\ \emph {et~al.}(2011)\citenamefont {Chen},
  \citenamefont {Gu},\ and\ \citenamefont {Wen}}]{Chen2011}%
  \BibitemOpen
  \bibfield  {author} {\bibinfo {author} {\bibfnamefont {X.}~\bibnamefont
  {Chen}}, \bibinfo {author} {\bibfnamefont {Z.-C.}\ \bibnamefont {Gu}},\ and\
  \bibinfo {author} {\bibfnamefont {X.-G.}\ \bibnamefont {Wen}},\ }\bibfield
  {title} {\bibinfo {title} {Complete classification of one-dimensional gapped
  quantum phases in interacting spin systems},\ }\bibfield  {journal} {\bibinfo
   {journal} {Physical Review B}\ }\textbf {\bibinfo {volume} {84}},\ \href
  {https://doi.org/10.1103/physrevb.84.235128} {10.1103/physrevb.84.235128}
  (\bibinfo {year} {2011})\BibitemShut {NoStop}%
\bibitem [{\citenamefont {Carleo}\ and\ \citenamefont
  {Troyer}(2017)}]{Carleo2017}%
  \BibitemOpen
  \bibfield  {author} {\bibinfo {author} {\bibfnamefont {G.}~\bibnamefont
  {Carleo}}\ and\ \bibinfo {author} {\bibfnamefont {M.}~\bibnamefont
  {Troyer}},\ }\bibfield  {title} {\bibinfo {title} {Solving the quantum
  many-body problem with artificial neural networks},\ }\href
  {https://doi.org/10.1126/science.aag2302} {\bibfield  {journal} {\bibinfo
  {journal} {Science}\ }\textbf {\bibinfo {volume} {355}},\ \bibinfo {pages}
  {602–606} (\bibinfo {year} {2017})}\BibitemShut {NoStop}%
\bibitem [{\citenamefont {Huang}\ \emph
  {et~al.}(2021{\natexlab{a}})\citenamefont {Huang}, \citenamefont {Kueng},
  \citenamefont {Torlai}, \citenamefont {Albert},\ and\ \citenamefont
  {Preskill}}]{Huang2021}%
  \BibitemOpen
  \bibfield  {author} {\bibinfo {author} {\bibfnamefont {H.-Y.}\ \bibnamefont
  {Huang}}, \bibinfo {author} {\bibfnamefont {R.}~\bibnamefont {Kueng}},
  \bibinfo {author} {\bibfnamefont {G.}~\bibnamefont {Torlai}}, \bibinfo
  {author} {\bibfnamefont {V.~V.}\ \bibnamefont {Albert}},\ and\ \bibinfo
  {author} {\bibfnamefont {J.}~\bibnamefont {Preskill}},\ }\href@noop {}
  {\bibinfo {title} {Provably efficient machine learning for quantum many-body
  problems}} (\bibinfo {year} {2021}{\natexlab{a}}),\ \Eprint
  {https://arxiv.org/abs/2106.12627} {arXiv:2106.12627 [quant-ph]} \BibitemShut
  {NoStop}%
\bibitem [{\citenamefont {Schuld}\ and\ \citenamefont
  {Petruccione}(2018)}]{Schuld2018}%
  \BibitemOpen
  \bibfield  {author} {\bibinfo {author} {\bibfnamefont {M.}~\bibnamefont
  {Schuld}}\ and\ \bibinfo {author} {\bibfnamefont {F.}~\bibnamefont
  {Petruccione}},\ }\href {https://doi.org/10.1007/978-3-319-96424-9} {\emph
  {\bibinfo {title} {Supervised Learning with Quantum Computers}}}\ (\bibinfo
  {year} {2018})\BibitemShut {NoStop}%
\bibitem [{\citenamefont {Mehta}\ \emph {et~al.}(2019)\citenamefont {Mehta},
  \citenamefont {Bukov}, \citenamefont {Wang}, \citenamefont {Day},
  \citenamefont {Richardson}, \citenamefont {Fisher},\ and\ \citenamefont
  {Schwab}}]{Mehta2019}%
  \BibitemOpen
  \bibfield  {author} {\bibinfo {author} {\bibfnamefont {P.}~\bibnamefont
  {Mehta}}, \bibinfo {author} {\bibfnamefont {M.}~\bibnamefont {Bukov}},
  \bibinfo {author} {\bibfnamefont {C.-H.}\ \bibnamefont {Wang}}, \bibinfo
  {author} {\bibfnamefont {A.~G.}\ \bibnamefont {Day}}, \bibinfo {author}
  {\bibfnamefont {C.}~\bibnamefont {Richardson}}, \bibinfo {author}
  {\bibfnamefont {C.~K.}\ \bibnamefont {Fisher}},\ and\ \bibinfo {author}
  {\bibfnamefont {D.~J.}\ \bibnamefont {Schwab}},\ }\bibfield  {title}
  {\bibinfo {title} {A high-bias, low-variance introduction to machine learning
  for physicists},\ }\href {https://doi.org/10.1016/j.physrep.2019.03.001}
  {\bibfield  {journal} {\bibinfo  {journal} {Physics Reports}\ }\textbf
  {\bibinfo {volume} {810}},\ \bibinfo {pages} {1–124} (\bibinfo {year}
  {2019})}\BibitemShut {NoStop}%
\bibitem [{\citenamefont {Mitchell}(1997)}]{Mitchell1997}%
  \BibitemOpen
  \bibfield  {author} {\bibinfo {author} {\bibfnamefont {T.~M.}\ \bibnamefont
  {Mitchell}},\ }\href@noop {} {\emph {\bibinfo {title} {Machine Learning}}}\
  (\bibinfo  {publisher} {McGraw-Hill Science, Engineering, Math},\ \bibinfo
  {address} {Cambridge, MS},\ \bibinfo {year} {1997})\BibitemShut {NoStop}%
\bibitem [{\citenamefont {Goodfellow}\ \emph {et~al.}(2016)\citenamefont
  {Goodfellow}, \citenamefont {Bengio},\ and\ \citenamefont
  {Courville}}]{Goodfellow2016}%
  \BibitemOpen
  \bibfield  {author} {\bibinfo {author} {\bibfnamefont {I.}~\bibnamefont
  {Goodfellow}}, \bibinfo {author} {\bibfnamefont {Y.}~\bibnamefont {Bengio}},\
  and\ \bibinfo {author} {\bibfnamefont {A.}~\bibnamefont {Courville}},\
  }\href@noop {} {\emph {\bibinfo {title} {Deep Learning}}}\ (\bibinfo
  {publisher} {The MIT Press},\ \bibinfo {address} {Cambridge,
  Massachussetts},\ \bibinfo {year} {2016})\BibitemShut {NoStop}%
\bibitem [{\citenamefont {Kamdar}\ \emph {et~al.}(2015)\citenamefont {Kamdar},
  \citenamefont {Turk},\ and\ \citenamefont {Brunner}}]{Kamdar2015}%
  \BibitemOpen
  \bibfield  {author} {\bibinfo {author} {\bibfnamefont {H.~M.}\ \bibnamefont
  {Kamdar}}, \bibinfo {author} {\bibfnamefont {M.~J.}\ \bibnamefont {Turk}},\
  and\ \bibinfo {author} {\bibfnamefont {R.~J.}\ \bibnamefont {Brunner}},\
  }\bibfield  {title} {\bibinfo {title} {Machine learning and cosmological
  simulations – i. semi-analytical models},\ }\href
  {https://doi.org/10.1093/mnras/stv2310} {\bibfield  {journal} {\bibinfo
  {journal} {Monthly Notices of the Royal Astronomical Society}\ }\textbf
  {\bibinfo {volume} {455}},\ \bibinfo {pages} {642–658} (\bibinfo {year}
  {2015})}\BibitemShut {NoStop}%
\bibitem [{\citenamefont {Kamdar}\ \emph {et~al.}(2016)\citenamefont {Kamdar},
  \citenamefont {Turk},\ and\ \citenamefont {Brunner}}]{Kamdar2016}%
  \BibitemOpen
  \bibfield  {author} {\bibinfo {author} {\bibfnamefont {H.~M.}\ \bibnamefont
  {Kamdar}}, \bibinfo {author} {\bibfnamefont {M.~J.}\ \bibnamefont {Turk}},\
  and\ \bibinfo {author} {\bibfnamefont {R.~J.}\ \bibnamefont {Brunner}},\
  }\bibfield  {title} {\bibinfo {title} {Machine learning and cosmological
  simulations – ii. hydrodynamical simulations},\ }\href
  {https://doi.org/10.1093/mnras/stv2981} {\bibfield  {journal} {\bibinfo
  {journal} {Monthly Notices of the Royal Astronomical Society}\ }\textbf
  {\bibinfo {volume} {457}},\ \bibinfo {pages} {1162–1179} (\bibinfo {year}
  {2016})}\BibitemShut {NoStop}%
\bibitem [{\citenamefont {Lochner}\ \emph {et~al.}(2016)\citenamefont
  {Lochner}, \citenamefont {McEwen}, \citenamefont {Peiris}, \citenamefont
  {Lahav},\ and\ \citenamefont {Winter}}]{Lochner2016}%
  \BibitemOpen
  \bibfield  {author} {\bibinfo {author} {\bibfnamefont {M.}~\bibnamefont
  {Lochner}}, \bibinfo {author} {\bibfnamefont {J.~D.}\ \bibnamefont {McEwen}},
  \bibinfo {author} {\bibfnamefont {H.~V.}\ \bibnamefont {Peiris}}, \bibinfo
  {author} {\bibfnamefont {O.}~\bibnamefont {Lahav}},\ and\ \bibinfo {author}
  {\bibfnamefont {M.~K.}\ \bibnamefont {Winter}},\ }\bibfield  {title}
  {\bibinfo {title} {Photometric supernova classification with machine
  learning},\ }\href {https://doi.org/10.3847/0067-0049/225/2/31} {\bibfield
  {journal} {\bibinfo  {journal} {The Astrophysical Journal Supplement Series}\
  }\textbf {\bibinfo {volume} {225}},\ \bibinfo {pages} {31} (\bibinfo {year}
  {2016})}\BibitemShut {NoStop}%
\bibitem [{\citenamefont {Torlai}\ \emph {et~al.}(2018)\citenamefont {Torlai},
  \citenamefont {Mazzola}, \citenamefont {Carrasquilla}, \citenamefont
  {Troyer}, \citenamefont {Melko},\ and\ \citenamefont {Carleo}}]{Torlai2018}%
  \BibitemOpen
  \bibfield  {author} {\bibinfo {author} {\bibfnamefont {G.}~\bibnamefont
  {Torlai}}, \bibinfo {author} {\bibfnamefont {G.}~\bibnamefont {Mazzola}},
  \bibinfo {author} {\bibfnamefont {J.}~\bibnamefont {Carrasquilla}}, \bibinfo
  {author} {\bibfnamefont {M.}~\bibnamefont {Troyer}}, \bibinfo {author}
  {\bibfnamefont {R.}~\bibnamefont {Melko}},\ and\ \bibinfo {author}
  {\bibfnamefont {G.}~\bibnamefont {Carleo}},\ }\bibfield  {title} {\bibinfo
  {title} {Neural-network quantum state tomography},\ }\href
  {https://doi.org/10.1038/s41567-018-0048-5} {\bibfield  {journal} {\bibinfo
  {journal} {Nature Physics}\ }\textbf {\bibinfo {volume} {14}},\ \bibinfo
  {pages} {447–450} (\bibinfo {year} {2018})}\BibitemShut {NoStop}%
\bibitem [{\citenamefont {Canabarro}\ \emph
  {et~al.}(2019{\natexlab{a}})\citenamefont {Canabarro}, \citenamefont
  {Brito},\ and\ \citenamefont {Chaves}}]{Canabarro2019a}%
  \BibitemOpen
  \bibfield  {author} {\bibinfo {author} {\bibfnamefont {A.}~\bibnamefont
  {Canabarro}}, \bibinfo {author} {\bibfnamefont {S.}~\bibnamefont {Brito}},\
  and\ \bibinfo {author} {\bibfnamefont {R.}~\bibnamefont {Chaves}},\
  }\bibfield  {title} {\bibinfo {title} {Machine learning nonlocal
  correlations},\ }\bibfield  {journal} {\bibinfo  {journal} {Physical Review
  Letters}\ }\textbf {\bibinfo {volume} {122}},\ \href
  {https://doi.org/10.1103/physrevlett.122.200401}
  {10.1103/physrevlett.122.200401} (\bibinfo {year}
  {2019}{\natexlab{a}})\BibitemShut {NoStop}%
\bibitem [{\citenamefont {Iten}\ \emph {et~al.}(2020)\citenamefont {Iten},
  \citenamefont {Metger}, \citenamefont {Wilming}, \citenamefont {del Rio},\
  and\ \citenamefont {Renner}}]{Iten2020}%
  \BibitemOpen
  \bibfield  {author} {\bibinfo {author} {\bibfnamefont {R.}~\bibnamefont
  {Iten}}, \bibinfo {author} {\bibfnamefont {T.}~\bibnamefont {Metger}},
  \bibinfo {author} {\bibfnamefont {H.}~\bibnamefont {Wilming}}, \bibinfo
  {author} {\bibfnamefont {L.}~\bibnamefont {del Rio}},\ and\ \bibinfo {author}
  {\bibfnamefont {R.}~\bibnamefont {Renner}},\ }\bibfield  {title} {\bibinfo
  {title} {Discovering physical concepts with neural networks},\ }\bibfield
  {journal} {\bibinfo  {journal} {Physical Review Letters}\ }\textbf {\bibinfo
  {volume} {124}},\ \href {https://doi.org/10.1103/physrevlett.124.010508}
  {10.1103/physrevlett.124.010508} (\bibinfo {year} {2020})\BibitemShut
  {NoStop}%
\bibitem [{\citenamefont {Carrasquilla}\ and\ \citenamefont
  {Melko}(2017)}]{Carrasquilla2017}%
  \BibitemOpen
  \bibfield  {author} {\bibinfo {author} {\bibfnamefont {J.}~\bibnamefont
  {Carrasquilla}}\ and\ \bibinfo {author} {\bibfnamefont {R.~G.}\ \bibnamefont
  {Melko}},\ }\bibfield  {title} {\bibinfo {title} {Machine learning phases of
  matter},\ }\href {https://doi.org/10.1038/nphys4035} {\bibfield  {journal}
  {\bibinfo  {journal} {Nature Physics}\ }\textbf {\bibinfo {volume} {13}},\
  \bibinfo {pages} {431} (\bibinfo {year} {2017})}\BibitemShut {NoStop}%
\bibitem [{\citenamefont {Dong}\ \emph {et~al.}(2019)\citenamefont {Dong},
  \citenamefont {Pollmann},\ and\ \citenamefont {Zhang}}]{Dong2019}%
  \BibitemOpen
  \bibfield  {author} {\bibinfo {author} {\bibfnamefont {X.-Y.}\ \bibnamefont
  {Dong}}, \bibinfo {author} {\bibfnamefont {F.}~\bibnamefont {Pollmann}},\
  and\ \bibinfo {author} {\bibfnamefont {X.-F.}\ \bibnamefont {Zhang}},\
  }\bibfield  {title} {\bibinfo {title} {Machine learning of quantum phase
  transitions},\ }\bibfield  {journal} {\bibinfo  {journal} {Physical Review
  B}\ }\textbf {\bibinfo {volume} {99}},\ \href
  {https://doi.org/10.1103/physrevb.99.121104} {10.1103/physrevb.99.121104}
  (\bibinfo {year} {2019})\BibitemShut {NoStop}%
\bibitem [{\citenamefont {Shiina}\ \emph {et~al.}(2020)\citenamefont {Shiina},
  \citenamefont {Mori}, \citenamefont {Okabe},\ and\ \citenamefont
  {Lee}}]{Shiina2020}%
  \BibitemOpen
  \bibfield  {author} {\bibinfo {author} {\bibfnamefont {K.}~\bibnamefont
  {Shiina}}, \bibinfo {author} {\bibfnamefont {H.}~\bibnamefont {Mori}},
  \bibinfo {author} {\bibfnamefont {Y.}~\bibnamefont {Okabe}},\ and\ \bibinfo
  {author} {\bibfnamefont {H.~K.}\ \bibnamefont {Lee}},\ }\bibfield  {title}
  {\bibinfo {title} {Machine-learning studies on spin models},\ }\href
  {https://doi.org/10.1038/s41598-020-58263-5} {\bibfield  {journal} {\bibinfo
  {journal} {Scientific Reports}\ }\textbf {\bibinfo {volume} {10}},\ \bibinfo
  {pages} {2177} (\bibinfo {year} {2020})}\BibitemShut {NoStop}%
\bibitem [{\citenamefont {Rem}\ \emph {et~al.}(2019)\citenamefont {Rem},
  \citenamefont {Käming}, \citenamefont {Tarnowski}, \citenamefont {Asteria},
  \citenamefont {Fläschner}, \citenamefont {Becker}, \citenamefont
  {Sengstock},\ and\ \citenamefont {Weitenberg}}]{Rem2019}%
  \BibitemOpen
  \bibfield  {author} {\bibinfo {author} {\bibfnamefont {B.~S.}\ \bibnamefont
  {Rem}}, \bibinfo {author} {\bibfnamefont {N.}~\bibnamefont {Käming}},
  \bibinfo {author} {\bibfnamefont {M.}~\bibnamefont {Tarnowski}}, \bibinfo
  {author} {\bibfnamefont {L.}~\bibnamefont {Asteria}}, \bibinfo {author}
  {\bibfnamefont {N.}~\bibnamefont {Fläschner}}, \bibinfo {author}
  {\bibfnamefont {C.}~\bibnamefont {Becker}}, \bibinfo {author} {\bibfnamefont
  {K.}~\bibnamefont {Sengstock}},\ and\ \bibinfo {author} {\bibfnamefont
  {C.}~\bibnamefont {Weitenberg}},\ }\bibfield  {title} {\bibinfo {title}
  {Identifying quantum phase transitions using artificial neural networks on
  experimental data},\ }\href {https://doi.org/10.1038/s41567-019-0554-0}
  {\bibfield  {journal} {\bibinfo  {journal} {Nature Physics}\ }\textbf
  {\bibinfo {volume} {15}},\ \bibinfo {pages} {917–920} (\bibinfo {year}
  {2019})}\BibitemShut {NoStop}%
\bibitem [{\citenamefont {Broecker}\ \emph {et~al.}(2017)\citenamefont
  {Broecker}, \citenamefont {Carrasquilla}, \citenamefont {Melko},\ and\
  \citenamefont {Trebst}}]{Broecker2017}%
  \BibitemOpen
  \bibfield  {author} {\bibinfo {author} {\bibfnamefont {P.}~\bibnamefont
  {Broecker}}, \bibinfo {author} {\bibfnamefont {J.}~\bibnamefont
  {Carrasquilla}}, \bibinfo {author} {\bibfnamefont {R.~G.}\ \bibnamefont
  {Melko}},\ and\ \bibinfo {author} {\bibfnamefont {S.}~\bibnamefont
  {Trebst}},\ }\bibfield  {title} {\bibinfo {title} {Machine learning quantum
  phases of matter beyond the fermion sign problem},\ }\bibfield  {journal}
  {\bibinfo  {journal} {Scientific Reports}\ }\textbf {\bibinfo {volume} {7}},\
  \href {https://doi.org/10.1038/s41598-017-09098-0}
  {10.1038/s41598-017-09098-0} (\bibinfo {year} {2017})\BibitemShut {NoStop}%
\bibitem [{\citenamefont {Canabarro}\ \emph
  {et~al.}(2019{\natexlab{b}})\citenamefont {Canabarro}, \citenamefont
  {Fanchini}, \citenamefont {Malvezzi}, \citenamefont {Pereira},\ and\
  \citenamefont {Chaves}}]{Canabarro2019}%
  \BibitemOpen
  \bibfield  {author} {\bibinfo {author} {\bibfnamefont {A.}~\bibnamefont
  {Canabarro}}, \bibinfo {author} {\bibfnamefont {F.~F.}\ \bibnamefont
  {Fanchini}}, \bibinfo {author} {\bibfnamefont {A.~L.}\ \bibnamefont
  {Malvezzi}}, \bibinfo {author} {\bibfnamefont {R.}~\bibnamefont {Pereira}},\
  and\ \bibinfo {author} {\bibfnamefont {R.}~\bibnamefont {Chaves}},\
  }\bibfield  {title} {\bibinfo {title} {Unveiling phase transitions with
  machine learning},\ }\href {https://doi.org/10.1103/PhysRevB.100.045129}
  {\bibfield  {journal} {\bibinfo  {journal} {Phys. Rev. B}\ }\textbf {\bibinfo
  {volume} {100}},\ \bibinfo {pages} {045129} (\bibinfo {year}
  {2019}{\natexlab{b}})}\BibitemShut {NoStop}%
\bibitem [{\citenamefont {Kitazawa}\ \emph {et~al.}(1996)\citenamefont
  {Kitazawa}, \citenamefont {Nomura},\ and\ \citenamefont
  {Okamoto}}]{Kitazawa1996}%
  \BibitemOpen
  \bibfield  {author} {\bibinfo {author} {\bibfnamefont {A.}~\bibnamefont
  {Kitazawa}}, \bibinfo {author} {\bibfnamefont {K.}~\bibnamefont {Nomura}},\
  and\ \bibinfo {author} {\bibfnamefont {K.}~\bibnamefont {Okamoto}},\
  }\bibfield  {title} {\bibinfo {title} {Phase diagram of
  $\mathit{S}\phantom{\rule{0ex}{0ex}}=\phantom{\rule{0ex}{0ex}}1$
  bond-alternating xxz chains},\ }\href
  {https://doi.org/10.1103/PhysRevLett.76.4038} {\bibfield  {journal} {\bibinfo
   {journal} {Phys. Rev. Lett.}\ }\textbf {\bibinfo {volume} {76}},\ \bibinfo
  {pages} {4038} (\bibinfo {year} {1996})}\BibitemShut {NoStop}%
\bibitem [{\citenamefont {Chen}\ \emph {et~al.}(2003)\citenamefont {Chen},
  \citenamefont {Hida},\ and\ \citenamefont {Sanctuary}}]{Chen2003}%
  \BibitemOpen
  \bibfield  {author} {\bibinfo {author} {\bibfnamefont {W.}~\bibnamefont
  {Chen}}, \bibinfo {author} {\bibfnamefont {K.}~\bibnamefont {Hida}},\ and\
  \bibinfo {author} {\bibfnamefont {B.~C.}\ \bibnamefont {Sanctuary}},\
  }\bibfield  {title} {\bibinfo {title} {Ground-state phase diagram of $s=1$
  $\mathrm{XXZ}$ chains with uniaxial single-ion-type anisotropy},\ }\href
  {https://doi.org/10.1103/PhysRevB.67.104401} {\bibfield  {journal} {\bibinfo
  {journal} {Phys. Rev. B}\ }\textbf {\bibinfo {volume} {67}},\ \bibinfo
  {pages} {104401} (\bibinfo {year} {2003})}\BibitemShut {NoStop}%
\bibitem [{\citenamefont {L{\"a}uchli}\ \emph {et~al.}(2006)\citenamefont
  {L{\"a}uchli}, \citenamefont {Schmid},\ and\ \citenamefont
  {Trebst}}]{Luchli2006}%
  \BibitemOpen
  \bibfield  {author} {\bibinfo {author} {\bibfnamefont {A.}~\bibnamefont
  {L{\"a}uchli}}, \bibinfo {author} {\bibfnamefont {G.}~\bibnamefont
  {Schmid}},\ and\ \bibinfo {author} {\bibfnamefont {S.}~\bibnamefont
  {Trebst}},\ }\bibfield  {title} {\bibinfo {title} {Spin nematics correlations
  in bilinear-biquadratic s=1 spin chains},\ }\bibfield  {journal} {\bibinfo
  {journal} {Physical Review B}\ }\textbf {\bibinfo {volume} {74}},\ \href
  {https://doi.org/10.1103/physrevb.74.144426} {10.1103/physrevb.74.144426}
  (\bibinfo {year} {2006})\BibitemShut {NoStop}%
\bibitem [{\citenamefont {Serneels}\ \emph {et~al.}(2006)\citenamefont
  {Serneels}, \citenamefont {De~Nolf},\ and\ \citenamefont
  {Van~Espen}}]{Serneels}%
  \BibitemOpen
  \bibfield  {author} {\bibinfo {author} {\bibfnamefont {S.}~\bibnamefont
  {Serneels}}, \bibinfo {author} {\bibfnamefont {E.}~\bibnamefont {De~Nolf}},\
  and\ \bibinfo {author} {\bibfnamefont {P.~J.}\ \bibnamefont {Van~Espen}},\
  }\bibfield  {title} {\bibinfo {title} {Spatial sign preprocessing: A simple
  way to impart moderate robustness to multivariate estimators},\ }\href
  {https://doi.org/10.1021/ci050498u} {\bibfield  {journal} {\bibinfo
  {journal} {Journal of Chemical Information and Modeling}\ }\textbf {\bibinfo
  {volume} {46}},\ \bibinfo {pages} {1402} (\bibinfo {year} {2006})},\ \bibinfo
  {note} {pMID: 16711760}\BibitemShut {NoStop}%
\bibitem [{\citenamefont {Müller}\ and\ \citenamefont
  {Guido}(2016)}]{Andreas2016}%
  \BibitemOpen
  \bibfield  {author} {\bibinfo {author} {\bibfnamefont {A.~C.}\ \bibnamefont
  {Müller}}\ and\ \bibinfo {author} {\bibfnamefont {S.}~\bibnamefont
  {Guido}},\ }\href@noop {} {\emph {\bibinfo {title} {Introduction to Machine
  Learning with Python}}}\ (\bibinfo  {publisher} {O’Reilly Media, Inc.},\
  \bibinfo {address} {Sebastopol, CA},\ \bibinfo {year} {2016})\BibitemShut
  {NoStop}%
\bibitem [{\citenamefont {Pedregosa}\ \emph {et~al.}(2011)\citenamefont
  {Pedregosa}, \citenamefont {Varoquaux}, \citenamefont {Gramfort},
  \citenamefont {Michel}, \citenamefont {Thirion}, \citenamefont {Grisel},
  \citenamefont {Blondel}, \citenamefont {Prettenhofer}, \citenamefont {Weiss},
  \citenamefont {Dubourg}, \citenamefont {Vanderplas}, \citenamefont {Passos},
  \citenamefont {Cournapeau}, \citenamefont {Brucher}, \citenamefont {Perrot},\
  and\ \citenamefont {Duchesnay}}]{scikit-learn}%
  \BibitemOpen
  \bibfield  {author} {\bibinfo {author} {\bibfnamefont {F.}~\bibnamefont
  {Pedregosa}}, \bibinfo {author} {\bibfnamefont {G.}~\bibnamefont
  {Varoquaux}}, \bibinfo {author} {\bibfnamefont {A.}~\bibnamefont {Gramfort}},
  \bibinfo {author} {\bibfnamefont {V.}~\bibnamefont {Michel}}, \bibinfo
  {author} {\bibfnamefont {B.}~\bibnamefont {Thirion}}, \bibinfo {author}
  {\bibfnamefont {O.}~\bibnamefont {Grisel}}, \bibinfo {author} {\bibfnamefont
  {M.}~\bibnamefont {Blondel}}, \bibinfo {author} {\bibfnamefont
  {P.}~\bibnamefont {Prettenhofer}}, \bibinfo {author} {\bibfnamefont
  {R.}~\bibnamefont {Weiss}}, \bibinfo {author} {\bibfnamefont
  {V.}~\bibnamefont {Dubourg}}, \bibinfo {author} {\bibfnamefont
  {J.}~\bibnamefont {Vanderplas}}, \bibinfo {author} {\bibfnamefont
  {A.}~\bibnamefont {Passos}}, \bibinfo {author} {\bibfnamefont
  {D.}~\bibnamefont {Cournapeau}}, \bibinfo {author} {\bibfnamefont
  {M.}~\bibnamefont {Brucher}}, \bibinfo {author} {\bibfnamefont
  {M.}~\bibnamefont {Perrot}},\ and\ \bibinfo {author} {\bibfnamefont
  {E.}~\bibnamefont {Duchesnay}},\ }\bibfield  {title} {\bibinfo {title}
  {Scikit-learn: Machine learning in {P}ython},\ }\href@noop {} {\bibfield
  {journal} {\bibinfo  {journal} {Journal of Machine Learning Research}\
  }\textbf {\bibinfo {volume} {12}},\ \bibinfo {pages} {2825} (\bibinfo {year}
  {2011})}\BibitemShut {NoStop}%
\bibitem [{\citenamefont {Fix}\ and\ \citenamefont
  {Hodges}(1951)}]{Evelyn1951}%
  \BibitemOpen
  \bibfield  {author} {\bibinfo {author} {\bibfnamefont {E.}~\bibnamefont
  {Fix}}\ and\ \bibinfo {author} {\bibfnamefont {J.~L.}\ \bibnamefont
  {Hodges}},\ }\bibfield  {title} {\bibinfo {title} {Discriminatory analysis,
  nonparametric discrimination: Consistency properties},\ }\href@noop {}
  {\bibfield  {journal} {\bibinfo  {journal} {USAF School of Aviation
  Medicine}\ } (\bibinfo {year} {1951})}\BibitemShut {NoStop}%
\bibitem [{\citenamefont {Huang}\ \emph
  {et~al.}(2021{\natexlab{b}})\citenamefont {Huang}, \citenamefont {Broughton},
  \citenamefont {Mohseni}, \citenamefont {Babbush}, \citenamefont {Boixo},
  \citenamefont {Neven},\ and\ \citenamefont {McClean}}]{Huang2021b}%
  \BibitemOpen
  \bibfield  {author} {\bibinfo {author} {\bibfnamefont {H.-Y.}\ \bibnamefont
  {Huang}}, \bibinfo {author} {\bibfnamefont {M.}~\bibnamefont {Broughton}},
  \bibinfo {author} {\bibfnamefont {M.}~\bibnamefont {Mohseni}}, \bibinfo
  {author} {\bibfnamefont {R.}~\bibnamefont {Babbush}}, \bibinfo {author}
  {\bibfnamefont {S.}~\bibnamefont {Boixo}}, \bibinfo {author} {\bibfnamefont
  {H.}~\bibnamefont {Neven}},\ and\ \bibinfo {author} {\bibfnamefont {J.~R.}\
  \bibnamefont {McClean}},\ }\bibfield  {title} {\bibinfo {title} {Power of
  data in quantum machine learning},\ }\href
  {https://doi.org/10.1126/science.aag2302} {\bibfield  {journal} {\bibinfo
  {journal} {Nat. Commun.}\ }\textbf {\bibinfo {volume} {12}},\ \bibinfo
  {pages} {1–9} (\bibinfo {year} {2021}{\natexlab{b}})}\BibitemShut {NoStop}%
\end{thebibliography}%
\end{document}